\documentclass[%
 reprint,
 amsmath,amssymb,
 aps,
 prx,
]{revtex4-2}

\usepackage{graphicx}
\usepackage{dcolumn}
\usepackage{bm}
\usepackage[dvipsnames]{xcolor}
\usepackage{lipsum}
\usepackage{mathrsfs}
\usepackage[colorlinks= true, linkcolor=blue, citecolor=blue, urlcolor=blue]{hyperref}

\UseRawInputEncoding

\begin{document}

\title{Reaction rates and the noisy saddle-node bifurcation: \\ Renormalization group for barrier crossing}

\author{David Hathcock}
\author{James P. Sethna}
\affiliation{Department of Physics,  Cornell University, Ithaca, New York 14853, USA}

\date{\today}

\begin{abstract}
Barrier crossing calculations in chemical reaction-rate theory typically assume that the barrier is large compared to the temperature. When the barrier vanishes, however, there is a qualitative change in behavior. Instead of crossing a barrier, particles slide down a sloping potential. We formulate a renormalization group description of this noisy saddle-node transition. We derive the universal scaling behavior and corrections to scaling for the mean escape time in overdamped systems with arbitrary barrier height. We also develop an accurate approximation to the full distribution of barrier escape times by approximating the eigenvalues of the Fokker-Plank operator as equally spaced. This lets us derive a family of distributions that captures the barrier crossing times for arbitrary barrier height. Our critical theory draws links between barrier crossing in chemistry, the renormalization group, and bifurcation theory. \end{abstract}

\pacs{Valid PACS appear here}
\maketitle

\section{Introduction}
In this paper, we investigate deep connections between barrier crossing, the renormalization group, and the noisy saddle node bifurcation. In particular, we show that Kramers' reaction rates can be understood as an asymptotic limit of the universal scaling near the continuous transition between high-barrier and barrier-less regimes. Applying methods from stochastic processes theory we derive an analytical expression for the universal scaling function for the mean barrier escape time near the critical point, giving the crossover between high and low barrier limits. The renormalization group provides a framework within which this result can be understood and systematically improved by perturbative calculations of corrections to scaling, some of which we give explicitly.

Barrier crossing arises in applications across physics, chemistry, and biology. In 1940, Kramers computed the barrier crossing rate for particles in both overdamped and underdamped regimes \cite{kramers1940brownian}. This result and others \cite{farkas1927keimbildungsgeschwindigkeit, becker1935kinetische, eyring1935activated} provided the theoretical explanation for the Arrhenius equation describing chemical rate coefficients $k\sim \exp(-E_b/k_B T)$, where $E_b$ is the energy barrier for activation \cite{arrhenius1889}. More recent efforts have established the escape rate at arbitrary damping, giving the crossover between the low- and high-damping limits \cite{mel1986theory, cartling1987kinetics}, and have accounted for the effects of state-dependent \cite{ermak1978brownian, lau2007statedependent}, non-gaussian \cite{bao2005cancellation, dybiec2007escape, baura2011colored}, and colored \cite{tsironis1988escape, hanggi1994escape, baura2011colored} noise, anharmonic corrections \cite{edholm1979accuracy, bez1981variational}, and fluctuating barriers \cite{bier1993matching, hanggi1994escape}.

Most transition-state calculations assume a large barrier limit. This means the barrier escape is a rare event, with a separation of time scales between the relaxation into a quasi-equilibrium state and the escape from that state \cite{hanggi1990reaction}. In the limit of vanishing barrier, however, there is a qualitative change in behavior. Particles instead slide down a monotonic potential, spending the most time near its inflection point. To capture the low barrier escape rate, extensions to Kramers' theory have been developed (e.g. incorporating anharmonic corrections), but these have significant errors when the barrier and thermal energy are comparable ($E_b \approx k_B T$) \cite{edholm1979accuracy}.

Finite barrier escape problems have garnered increasing theoretical interest over the past decade, with several studies contributing further low barrier refinements of existing theories \cite{palyulin2012finite, mazo2013thermal, aktaev2014theoretical, pollak2014finite, pollak2016kramers, bai2018simple, mazo2013thermal} or focusing directly on the saddle-node bifurcation where the barrier vanishes \cite{miller2012escape, herbert2017predictability}. Such escape processes are relevant to certain high precision measurements. For instance, force spectroscopy experiments apply a force on a single bond in a biomolecule until it breaks \cite{husson2009force, mazo2013thermal}. For typical molecules, the critical force, at which the energy barrier for breaking vanishes and Kramers' theory breaks down, is now well within the reach of atomic force microscopy and optical tweezers \cite{husson2009force}. Another exciting application is in micro- and nano-electromechanical devices, which sensitively switch oscillation amplitude in response to an input signal by operating near the barrier-less critical point \cite{miller2012escape, tadokoro2018driven}. Here, an analytical theory of low barrier crossing would help to distinguish between noise and signal activated switching.

We develop a critical theory for barrier crossing with a renormalization group approach that gives a complete scaling description of the noisy saddle-node bifurcation.
We are inspired by previous work on the `intermittency' 
	\footnote{In this paper `intermittency' refers to the chaotic dynamical intermittency studied in Refs.~\cite{hirsch1982intermittency, hirsch1982intermittencyRG, hu1982exact} that emerges in discrete maps via a tangency bifurcation. We use this term sparingly to avoid confusion with intermittency in fluid dynamics and other areas.}
route to chaos \cite{hirsch1982intermittency, hirsch1982intermittencyRG, hu1982exact}, where the renormalization group coarse-grains in time, then rescales the system to fix a certain term in the potential. In chaos theory, this procedure involves iterating and rescaling a discrete map \cite{hirsch1982intermittencyRG, hu1982exact}, leading to a different fixed point for the same renormalization group equations used by Feigenbaum to study period doubling \cite{feigenbaum1978quantitative}. We take the continuous time limit, reducing the renormalization group to a series of elementary rescalings and yielding a simplified description applicable to barrier escape problems. Our procedure organizes what amounts to dimensional analysis, providing an elegant renormalization-group framework that unifies Kramers' theory for Arrhenius barrier crossing with the dynamical systems theory of a noisy saddle-node bifurcation. 

Why do we frame our analysis in terms of the renormalization group, if 
the scaling form can be justified using dimensional analysis and the 
analytical methods we use are drawn from more traditional stochastic analysis?
On the one hand, this forms a wonderful case study, unifying and
illuminating bifurcation theory, the renormalization group, and chemical
reaction theory. Second, barrier crossing forms the solvable limiting case
of much more complex phenomena: coupling to colored-noise heat baths,
nucleation of abrupt phase transitions, and depinning transitions in disordered
systems (see Section~\ref{discussionSection}).
The scale invariance of random walks does not demand a renormalization
group proof of the central limit theorem, and the one-dimensional Ising
model can be solved without the machinery of flows in Hamiltonian space. 
But framing the problems in terms of the renormalization group provide
excellent pedagogical exercises, and the natural framework for extensions to
self-avoiding random walks and Ising models in higher dimensions.

Using our scaling theory as an organizing framework, we derive an analytical expression for the scaling form of the mean escape time near the saddle node bifurcation. We also compute corrections to scaling due to anharmonicity in the potential and finite initial or final position. Going beyond the mean, we develop an accurate approximation to the scaling form for the full distribution of escape times by assuming the eigenvalues of the Fokker-Plank operator are equally spaced.

As a starting point, we consider the equation of motion for an overdamped particle in a general potential $V(x)$ and driven by spatially dependent white noise,
\begin{equation}\label{overDampedLimit}
\dot{x} = f(x) + g(x) \, \xi (t).
\end{equation}
Here $f(x) = -\eta^{-1} \, dV/dx$ is the force exerted on the particle (divided by the damping coefficient $\eta$) and $g(x)$ is the spatially varying noise amplitude (with the damping absorbed). The noise $\xi(t)$ has zero mean, $\langle \xi(t) \rangle = 0$ and is uncorrelated in time, $ \langle \xi(t) \xi(t') \rangle = \delta(t-t')$. With barrier crossing phenomena in mind, we consider potentials with boundary conditions $V(x) \rightarrow \infty$ as $x\rightarrow -\infty$ and $V(x) \rightarrow -\infty$ as $x \rightarrow \infty$. The potential either has a single barrier or is monotonically decreasing (e.g. Figure~\ref{fig:barrierDiagram}). The quantity of interest is the mean barrier crossing time $\tau$, defined as the time particles take to reach $+\infty$ from an initial position at $-\infty$.

Besides the experimental systems discussed above, this model also serves as the natural description for a general chemical reaction, involving the transition between metastable species $A$ and $B$. These species are points in a $3N$ dimensional configuration space defined by the locations of $N$ reaction constituents. As derived by H\"{a}nggi \emph{et al.}  this system can be reduced to the one-dimensional model we study \cite{hanggi1990reaction}. They coarse-grain to a one dimensional reaction coordinate, which parameterizes the minimal gradient path between the states $A$ and $B$, neglecting effects of memory friction and noise correlations, and taking the overdamped limit produces Eq.~(\ref{overDampedLimit}). The effective potential along the reaction coordinate has a barrier separating species $A$ and $B$.

\section{Renormalization Group and Scaling Theory}\label{scalingSection}

\begin{figure}[t]
\includegraphics[width=\linewidth]{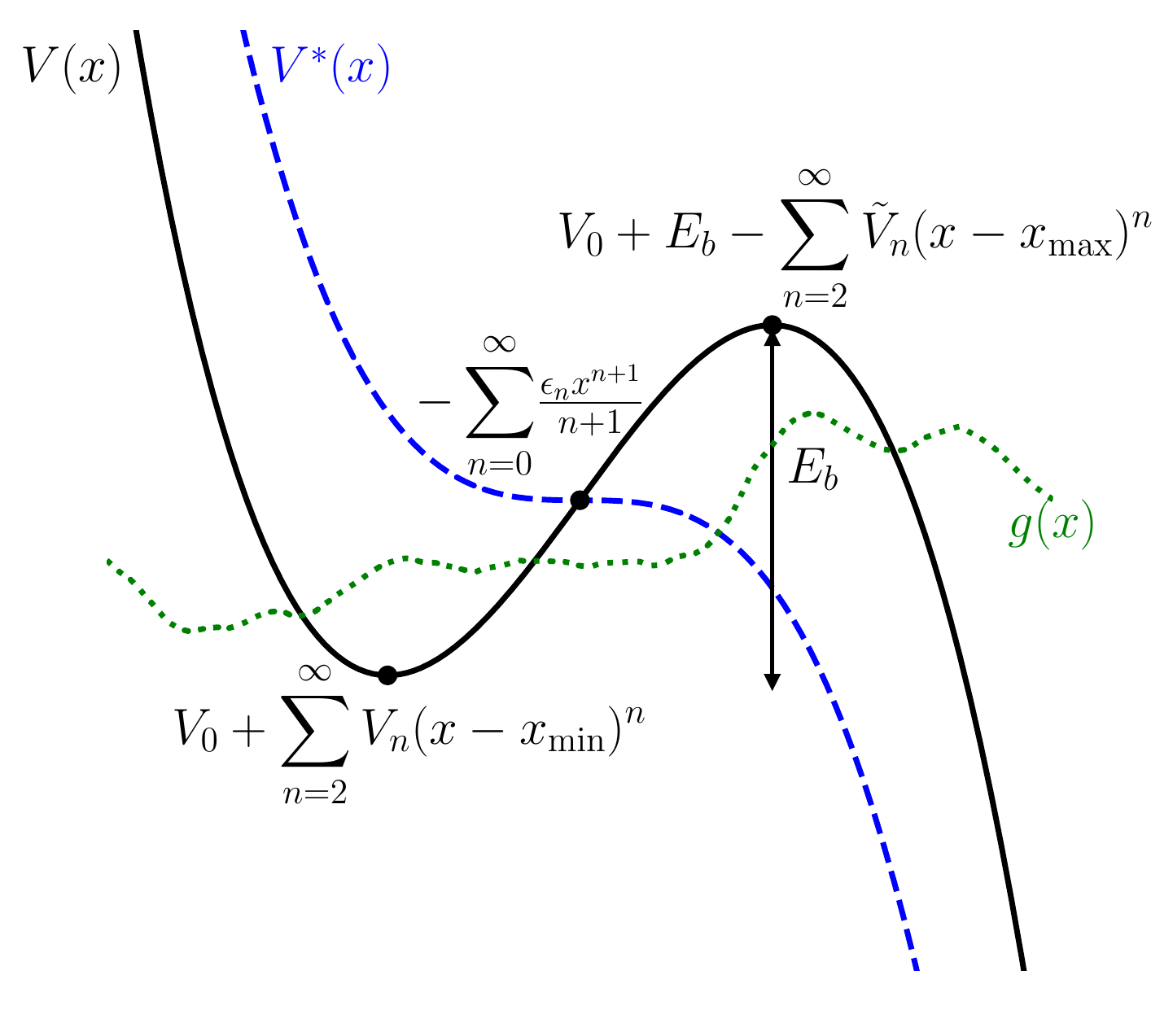}
\caption{\label{fig:barrierDiagram} Typical potentials in the high barrier Arrhenius limit (solid curve) and at the renormalization group fixed point (dashed curve). Kramers' theory utilizes a two point series expansion at $x_\text{min}$ in the potential well and at $x_\text{max}$, the top of the barrier. For our renormalization group approach the natural description is in terms of a single expansion at the origin parameterizing perturbations away from the fixed point potential $V^*(x) \propto - x^3$. Also shown is the noise amplitude g(x), which generically has spatial dependence (dotted curve).}
\end{figure}

To begin, we parameterize Eq.~(\ref{overDampedLimit}) by the Taylor coefficients of $g(x)$ and $f(x)$,
\begin{equation}\label{EOMseries}
\frac{d x}{d t} = \sum_{n = 0}^\infty \epsilon_n x^n + \xi(t) \sum_{n=0}^\infty g_n x^n.
\end{equation}
The renormalization group defines a flow in this space of systems described by a single reaction coordinate $x$. Near the renormalization group fixed point, the behavior is most effectively described by a single Taylor expansion at the origin. In contrast, for large barriers in Kramers' theory, the escape time is characterized by two expansions, capturing the harmonic oscillations in the potential well and at the top of the barrier. These two equivalent schemes are shown in Fig. \ref{fig:barrierDiagram}. Given the later expansion at the two extrema, the expansion at the origin can be reconstructed via a two-point Pad\'{e} approximation \cite{bender1999advanced}.

As discussed above, the discrete renormalization group coarse-grains by iterating a map, evolving the equations forward in time. Ignoring the noise for the moment, we consider a discrete approximation to Eq.~(\ref{overDampedLimit}), $x_{n+1} = x_n + \delta t f(x_n) \equiv h(x_n)$. References \cite{hirsch1982intermittency, hirsch1982intermittencyRG, hu1982exact} study this discrete equation using the Feigenbaum renormalization group transformation. This transformation iterates the map and rescales space, inducing a flow in function space, $T[h](x) = a h(h(x/a))$, where $a$ is the rescaling factor. The renormalization fixed point is the function $h(x)$ that obeys $T[h]=h$ for a particular rescaling $a$. We simplify this calculation by taking a continuum limit. Expanding the renormalization group to first order in $\delta t$, $T[h](x) = x +2 \delta t  (a f(x/a))$. Thus, in the continuous time limit, the RG iteration becomes a simple rescaling of time and space. We will use this below to derive our scaling theory for barrier crossing near a saddle node bifurcation.

Within the context of the renormalization group for singular perturbations developed by Goldenfeld, Oono, and others \cite{goldenfeld1989intermediate, goldenfeld1990anomalous}, our problem can be understood as having zero anomalous dimension. For our calculations in the following sections we do not need to use the traditional renormalization-group machinery; instead we assemble a variety of tools from probability theory and Markov processes, to express the mean and distributions of escape times as universal scaling functions near the transition where the barrier vanishes. The scaling theory provides a powerful and elegant structure which organizes our understanding barrier crossing.

Following the above expansion of the discrete renormalization group, we `coarse grain' the system in time by scaling, $\hat{t} = t/b$. As the time-scale shrinks, the noise is amplified, $\hat{\xi}(\hat t) = b^{1/2} \xi(t)$ (the exponent 1/2 follows from the units of the correlation function).
Our goal is to understand the scaling properties near the critical point, where a qualitative change in behavior occurs. For a generic analytic potential this happens when the barrier vanishes and $V(x) = -x^3$ is locally a perfect cubic. Therefore, we rescale our system to fix the coefficient $\epsilon_2$, corresponding to the cubic term in the potential. The correct rescaling defines a new spatial coordinate $\hat{x} = b x$. After both coarse graining and rescaling, we arrive at 
\begin{equation}
\frac{d \hat{x}}{d \hat{t}} = \sum_{n = 0}^\infty \epsilon_n b^{2-n} \hat{x}^n + \hat{\xi}(\hat t) \sum_{n=0}^\infty b^{3/2-n} g_n \hat{x}^n.
\end{equation}
We can then read off how the parameters flow under the renormalization group, $\hat{\epsilon}_n = b^{2-n} \epsilon_n$ and $\hat{g}_n = b^{3/2-n} g_n$. These flows and exponents exactly match those found under the discrete-time renormalization group \cite{hirsch1982intermittencyRG, hu1982exact}, indicating that the scaling of the 'intermittency route to chaos' \cite{Note1} is also non-anomalous \cite{goldenfeld1990anomalous}. Taking the coarse graining factor to be close to 1, $b = (1+ d\ell)$, we obtain continuous flow equations,
\begin{equation}
\frac{d\epsilon_n}{d\ell} = (2-n) \epsilon_n, \quad \frac{dg_n}{d\ell} = (3/2-n) g_n.
\end{equation}

The eigenfunctions of the renormalization group in our continuum theory are the monomials $x^n$ and  noisy monomials $\xi(t) x^n$. If the right hand side of Eq.~(\ref{EOMseries}) is an eigenfunction, it is scaled by a constant factor under the action of the renormalization group. These eigenfunctions are the much simpler continuous time limit of those for the discrete-time renormalization group \cite{hu1982exact}. In particular, the cubic potential $V(x) \propto -x^3$ (without noise) is the fixed point. At the fixed point, particle trajectories $x(t) \sim 1/t$ exhibit scale invariance in time as they approach the cubic inflection point at $x=0$. Perturbations away from the fixed point lead to dynamics with non-power law decay to a locally stable state or over the inflection point.

The mean barrier crossing time is a function of the potential shape and the noise correlation, encoded through the expansion coefficients $\epsilon_n$ and $g_n$. Thus, the escape time can be expressed as $\tau(\{\epsilon_n\}, \{g_n\})$, where $n\geq0$. If we coarse-grain until $g_0(\ell^*) = 1$, we find that the escape time has the form
\begin{equation}\label{generalScalingForm}
\tau =g_0^{-2/3} \mathcal{T} \left( \{\epsilon_n/g_0^{2(2-n)/3}\} , \{g_n/g_0^{1-2n/3} \} \right),
\end{equation} 
where $\mathcal{T}$ is a universal scaling function, with $n\geq1$ for the second term in brackets.

While the scaling form Eq.~(\ref{generalScalingForm}) could have been written down using dimensional analysis, the renormalization group approach provides the natural structure and motivation for our approach. The parameter space flows indicate that, with a fixed quadratic force, the constant and linear force and noise terms $\{\epsilon_0, \epsilon_1, g_0, g_1\}$ are relevant, growing under coarse graining and dominant on long time scales. Other variables are irrelevant and can be incorporated perturbatively. Of the relevant variables, the linear force coefficient $\epsilon_1$ can be set to zero by placing the origin at the inflection point of the potential. The spatial dependence of the noise (including the relevant linear term $g_1$) can also be removed by a change of coordinates $x \rightarrow \tilde{x}$ with $\tilde{x}$ defined by \cite{stratonovich1967topics}
\begin{equation}
x = \int^{\tilde{x}} \frac{g_0}{g(y)} \, dy,
\end{equation}
producing a system with constant noise $\tilde{g}(\tilde x) = g_0$ and force $\tilde{f}(\tilde{x}) = f(\tilde{x})/g(\tilde{x})$ (hence $g_1$ was relevant before we removed it, because it contributes to the linear term in the expansion of $\tilde{f}$).

Systems near enough to the critical point therefore can be modeled as a cubic potential with a linear perturbation $V(x) = -x^3/3 - \epsilon_0 x$ and constant noise $g_0$. This is the `normal form' used in bifurcation theory for the saddle node transition, and might have been anticipated from Taylor's theorem. The escape time scaling form becomes,
\begin{equation}
\tau =g_0^{-2/3} \mathcal{T} (\epsilon_0/g_0^{4/3}).
\end{equation}
Thus, the problem asymptotically reduces to finding the universal function of a single variable $\mathcal{T}(\alpha)$, where $\alpha = \epsilon_0/g_0^{4/3}$. The limiting form of the scaling function $\mathcal{T}(\alpha)$ must give the known solutions. In the limit $\alpha \rightarrow  -\infty$ the barrier is large compared to the noise, so the Kramers approximation  \cite{kramers1940brownian} applies, so
\begin{equation}\label{kramersLimit}
\mathcal{T}(\alpha) \sim \frac{\pi}{|\alpha|^{1/2}} e^{\frac{8}{3} |\alpha|^{3/2}}, \quad \alpha \to -\infty.
\end{equation}
For our choice of parameters, the energy barrier is given by $E_b/k_B T = 8/3 |\alpha|^{3/2}$. In the opposite limit $\alpha \rightarrow \infty$, the potential is downward sloping with gradient much larger than the noise level. The passage of particles over the inflection point occurs even in the absence of noise (in contrast to the Kramers limit, which requires noise for barrier escape). Therefore, the crossing time approaches that for a deterministic particle in the cubic potential. One can easily show that the limiting scaling form is 
\begin{equation}\label{deterministicLimit}
\mathcal{T}(\alpha) \sim \frac{\pi}{\alpha^{1/2}}, \quad \alpha \to \infty.
\end{equation}

\section{Mean Escape Time}
\subsection{Analytical escape time for relevant variables}\label{meanTime}

We now turn our focus to obtaining an exact analytical expression for $\mathcal{T}(\alpha)$ that is valid for all $\alpha$. To this end, we study the trajectories of particles injected at position $x_i$ and time $t_i$  into a general potential $V(x)$ with noise $g_0$ and compute the mean first passage time to $x_f$, following the standard approach \cite{hanggi1990reaction, malakhov2002evolution, hirsch1982intermittency}. Let $\mathcal{P}(x, t)$ be the distribution of particles over positions $x$ at time $t$, with $\mathcal{P}(x, t_i) = \delta(x-x_i)$. The probability that a particle has not reached $x_f$ at time $t$ is 
\begin{equation}\label{survivalProb}
\mathcal{P}(t) = \int_{-\infty}^{x_f} \mathcal{P}(x, t)\,  dx.
\end{equation}
Note that $\mathcal{P}(0) = 1$ and $\mathcal{P}(t) \rightarrow 0$ as $t \rightarrow \infty$ as long as there is noise driving the system, which guarantees particles reach $x_f$. The distribution of first passage times is $p(t) = -d\mathcal{P}/dt$ so that the mean first passage time is
\begin{equation}\label{mfpt}
\tau(x_i | x_f) = \int_0^\infty  t \, p(t)\,  dt = \int_0^\infty \mathcal{P}(t) \, dt,
\end{equation}
where we integrate by parts for the second equality. To derive a differential equation for $\tau(x_i | x_f)$, we start from the Kolmogorov backward equation for distribution $\mathcal{P}(x, t)$ with initial condition $x_i$ \cite{stratonovich1963topics}, 
\begin{equation}
-\frac{d\mathcal{P}(x, t) }{dt_i}= -V'(x_i) \frac{d \mathcal{P}(x, t)}{d x_i} + \frac{1}{2} g_0^2 \frac{d^2 \mathcal{P}(x,t)}{dx_i^2}.
\end{equation}
To write this equation in terms of the mean first passage time $\tau$, we multiply both sides by $t$ and integrate over $x$ and $t$. Using the relations in Eqs.~(\ref{survivalProb}) and (\ref{mfpt}) and the identity $d\mathcal{P}(x,t)/dt_i = -d \mathcal{P}(x,t)/dt$, we arrive at
\begin{equation}\label{fptODE}
\frac{1}{2} g_0^2 \tau''(x_i | x_f) - V'(x_i) \tau'(x_i| x_f)  = -1.
\end{equation}
This gives an ordinary differential equation for the first passage time from $x_i$ to $x_f$ of particles in potential $V(x)$ and constant noise with amplitude $g_0$. The boundary conditions are $\tau(x_f|x_f) = 0$ and $\tau'(-\infty|x_f) = 0$, which encode absorbing and reflecting boundaries respectively. Writing the solution to Eq.~(\ref{fptODE}) in integral form, we arrive at the result obtained in Refs. \cite{hanggi1990reaction, malakhov2002evolution, hirsch1982intermittency},
\begin{equation}\label{fptFunctional}
\tau(x_i | x_f)  = \frac{2}{g_0^2}\int_{x_i}^{x_f} dy \int_{-\infty}^y dz \, e^{-\frac{2}{g_0^2} \left[V(z) - V(y) \right]},
\end{equation}
which satisfies the boundary conditions as long as $V'(x) \rightarrow  \infty$ as $x\rightarrow - \infty$. For large barriers, it is known that Eq.~(\ref{fptFunctional}) reproduces Kramers escape rate formula via a saddle point approximation that expands the potential around the maximum and the minimum (as shown in Fig.~\ref{fig:barrierDiagram}) to second order \cite{hanggi1990reaction}.

\begin{figure}[t]
\includegraphics[width=\linewidth]{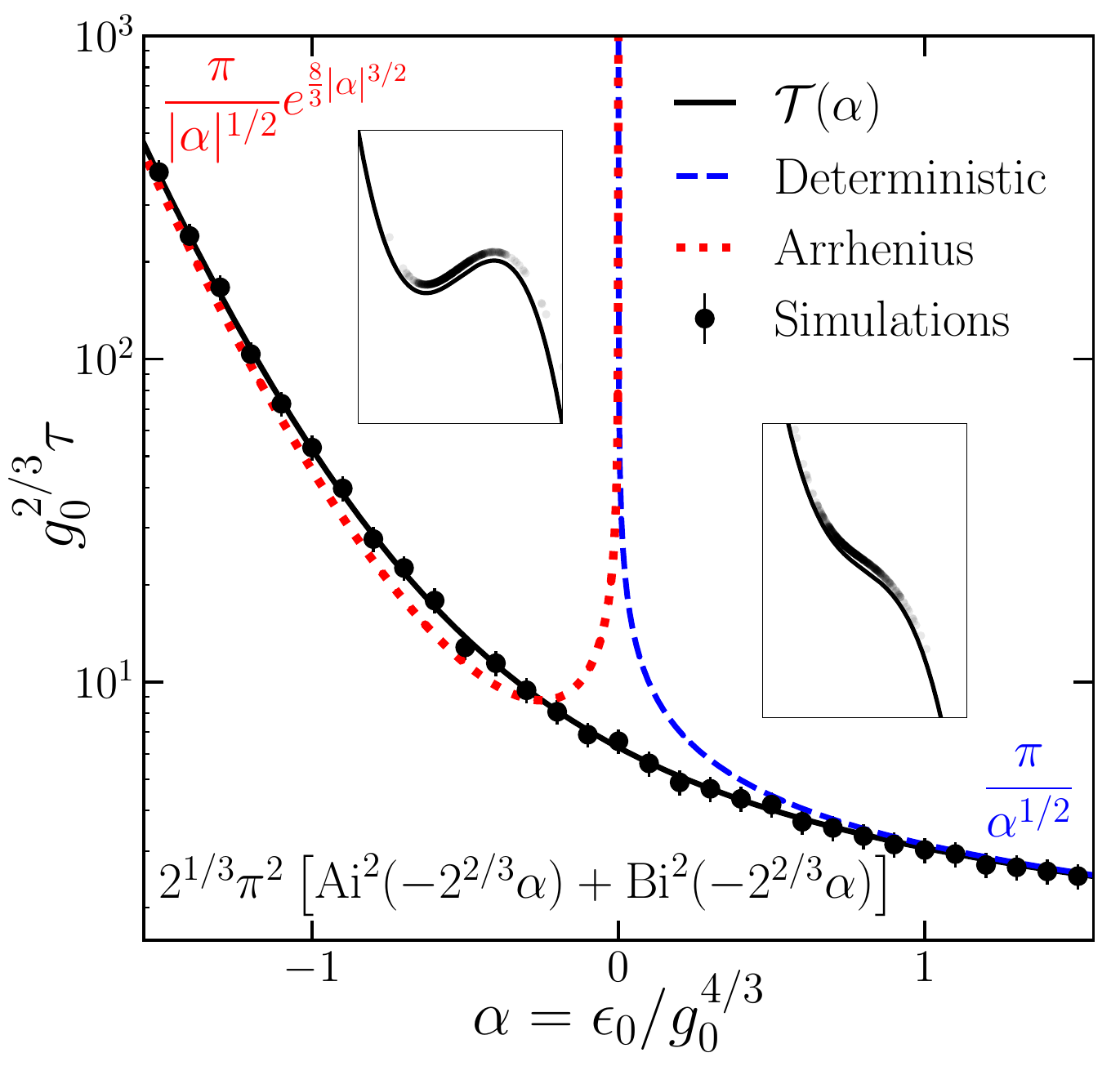}
\caption{\label{fig:escapeScaling} Comparison of the universal scaling function $\mathcal{T}(\alpha)$ (solid curve) to the Arrhenius (dotted curve) and deterministic (dashed curve) limits. Also shown are the mean escape times for 500 simulations of the barrier escape process. For the simulations we fixed $g_0=1$ while varying $\epsilon_0$ and used boundary conditions $x_f = - x_i = 25$. Agreement with our analytic expression for $\mathcal{T}(\alpha)$ is excellent. The insets show snapshots of the barrier crossing simulations for $\epsilon_0 = \pm 1$.}
\end{figure}

Our renormalization group analysis allows us to restrict our focus to the relevant variables. For the cubic potential (systems on the unstable manifold of the renormalization group fixed point), the escape time can be computed analytically using Eq.~(\ref{fptFunctional}) in the limit $x_f = -x_i \rightarrow \infty$. We find that $\tau = g_0^{-2/3} \mathcal{T}(\alpha)$ with the universal scaling function given by
\begin{equation}\label{exactSolution}
\mathcal{T}(\alpha) = 2^{1/3} \pi^2 \left[ \text{Ai}^2(-2^{2/3} \alpha ) + \text{Bi}^2( -2^{2/3} \alpha) \right], 
\end{equation}
where $\text{Ai}(x)$ and $\text{Bi}(x)$ are the first and second Airy functions and $\alpha = \epsilon_0/g_0^{4/3}$ as above. This solution is shown in Figure~\ref{fig:escapeScaling}, along with the Arrhenius and deterministic limits given in Eqs.~(\ref{kramersLimit}) and (\ref{deterministicLimit}) respectively and the mean barrier crossing times from direct simulations of the Langevin process [Eq.~(\ref{overDampedLimit})]. The universal scaling function $\mathcal{T}(\alpha)$ reproduces the two known limits when the barrier is large or the potential is strongly downward sloping and agrees excellently with the numerical results. 

Kramers' escape rate for the cubic potential follows from Eq.~(\ref{exactSolution}) and the asymptotic form of the second Airy function. As $\alpha \rightarrow 0$, however, contributions from the first Airy function become important so that Kramers' theory and extensions involving anharmonic corrections break down. The difference between Eqs.~(\ref{kramersLimit}) and (\ref{exactSolution}) is also related to the narrowing of the spectral gap of the barrier crossing Fokker-Plank operator (which has been measured numerically \cite{zhan2019diffusion} and is discussed below in Section~\ref{distributionSection}).

\subsection{Corrections to scaling}

\subsubsection{Finite launching and absorbing positions}
One limitation of our result Eq.~(\ref{exactSolution}) is that we assume initial and final states at infinity. In a real chemical or mechanical system, the transition of interest generally occurs between states with finite coordinates. For systems with large barriers, the escape time is exponentially large compared to the time it takes to settle into a metastable state in the well. Thus, the scaling function Eq.~(\ref{exactSolution}) is universal: independent of initial and final conditions. When the potential is downward sloping, the barrier crossing time is still dominated by the time spent near the inflection point and we can systematically compute corrections to scaling due to the finite initial and final positions.

Working in terms of scaling variables $\alpha = \epsilon_/g_0^{4/3}$, $\chi_- = x_i/g_0^{2/3}$, and $\chi_+ = x_f/g_0^{2/3}$, we can write the barrier crossing time for the cubic potential with arbitrary initial and final conditions as $\tau = g_0^{-2/3} \mathcal{T}(\alpha, \chi_-, \chi_+)$ with the scaling function,
\begin{equation}
\mathcal{T}(\alpha, \chi_-, \chi_+) = \mathcal{T}(\alpha) - \mathcal{T}_-(\alpha, \chi_-) - \mathcal{T}_+(\alpha, \chi_+).
\end{equation}
Here $\mathcal{T}_\pm(\alpha, \chi_\pm)$ are the universal corrections for finite final and initial conditions respectively. These have integral representations, 
\begin{equation}\label{fptCorrsectionInit}
\mathcal{T}_-(\alpha, \chi_-)  = 2 \int_{-\infty}^{\chi_-} dy \int_{-\infty}^y dz \, e^{- 2\left(y^3/3 + \alpha y -z^3/3 - \alpha z\right)}
\end{equation}
and 
\begin{equation}\label{fptCorrectionFinal}
\mathcal{T}_+(\alpha, \chi_+)  = 2 \int_{\chi_+}^{\infty} dy \int_{-\infty}^y dz \, e^{- 2\left(y^3/3 + \alpha y -z^3/3 - \alpha z\right)}.
\end{equation}
Assuming $\chi_-<0$ and $\chi_+>0$ these can be expanded in powers of $1/\chi_\pm$, 
\begin{equation}\label{fptCorrectionPerturb}
\begin{split}
\mathcal{T}_\pm(\alpha, \chi_\pm)  = & |\alpha|^{-1/2} \tan^{-1}(|\alpha|^{1/2} /|\chi_\pm|) \\
&\quad + \frac{1}{\chi_\pm^4} \left(\frac{1}{4} 
-\frac{\alpha}{2 \chi_\pm^2} + \frac{3\alpha^2}{4 \chi_\pm^4}\right) \\
 &\quad \quad \pm \frac{1}{\chi_\pm^7}\left(\frac{5}{14} -\frac{13 \alpha}{9 \chi_\pm^2}\right) + O(\chi_\pm^{-10}).
\end{split}
\end{equation}
The first term, which we can compute exactly, is simply the correction due to the deterministic trajectory in the cubic potential between $\chi_\pm$ and $\pm \infty$. The higher order terms describe the influence of noise on the corrections due to finite initial conditions. These terms appear to have the form $\chi_\pm^{-1-3j} f_j(\alpha/\chi_\pm^2)$ for integers $j$ and some functions $f_j$. The expansions of $f_1$ and $f_2$ are given in parenthesis in the expression above.

Our corrections to scaling for finite launching and absorbing positions are universal if the initial conditions are sufficiently close to the cubic inflection point so that anharmonic corrections are small. Thus, for an arbitrary potential, the universal escape time near a saddle node bifurcation is the given by Eq.~(\ref{exactSolution}) corrected using Eq.~(\ref{fptCorrectionPerturb}). In the following section we will separately treat the anharmonic corrections to scaling. The interplay between corrections to scaling due to anharmonicity and initial conditions will be an interesting subject for future studies.

\subsubsection{Anharmonic corrections}

The scaling function Eq.~(\ref{exactSolution}) also serves as a starting point from which the theory can be systematically improved by computing anharmonic corrections to scaling. The higher order terms in the potential are irrelevant variables under the renormalization group flows and hence can be treated perturbatively. For instance, consider a quartic perturbation $\delta V(x) = -\epsilon_3 x^4/4$ and let $\beta = \epsilon_3 g_0^{2/3}$. In the Kramers regime, $\alpha \rightarrow -\infty$, we have that $\mathcal{T}(\alpha, \beta) \approx \mathcal{T}(\alpha) + \beta^2 \mathcal{T}_3(\alpha)$ to leading order, where
\begin{equation}\label{t3Less}
\mathcal{T}_3(\alpha) \xrightarrow[]{\alpha \ll 0} \pi \sqrt{|\alpha|}  e^{\frac{8}{3} |\alpha|^{3/2}} (8 |\alpha|^{3/2}+11)/8.
\end{equation}
In the deterministic regime, $\alpha \rightarrow \infty$, we also add a quintic term as a regulator on the boundary conditions of the potential, $\delta V(x) = -\epsilon_3 x^4/4 - \epsilon_4 x^5/5$, with $\epsilon_4>0$ and sufficiently large so that the potential remains monotonically decreasing. To quadratic order in $\beta$ and $\gamma = \epsilon_4 g_0^{4/3}$, the universal scaling function is %
		\footnote{Here and in Eq.~\ref{t3Less} we could have avoided branch cuts in our scaling functions by using the variables $\sqrt{\alpha}$, $\sqrt{\beta}$, and $\sqrt{\gamma}$. We choose our convention to avoid complex valued scaling variables and imaginary numbers in Eq.~\ref{t3Less}.}
\begin{eqnarray}
 \mathcal{T}(\alpha, \beta, \gamma)& &\xrightarrow[]{\alpha \gg 0} \dfrac{\pi}{\sqrt{\alpha}} - \beta^2 \left( \dfrac{15}{8} \pi \sqrt{\alpha} - \dfrac{3 \pi}{4 \sqrt{\gamma}} \right) \\
&  &  \,  - \pi \sqrt{\gamma} + \dfrac{3}{2} \pi \sqrt{\alpha} \gamma  - \dfrac{5}{2} \pi \alpha \gamma^{3/2} + \dfrac{35}{8} \pi \alpha^{3/2} \gamma^{2}. \nonumber
\end{eqnarray}
The term $\pi/\sqrt{\alpha}$ is just the deterministic limit of the scaling form for the cubic potential and $\beta^2 \mathcal{T}_3(\alpha) = -15 \pi \sqrt{\alpha} \beta^2/8$ comes from the quartic perturbation to the inflection point. Other terms arise from quintic corrections or global changes in the potential. Here $\gamma$ is a dangerous irrelevant variable \cite[Sections 3.6, 5.4, \& 5.6]{cardy1996scaling}, which has a pole $3 \pi \beta^2/4\sqrt{\gamma}$ in the expansion about 0, because it is needed to keep the potential monotonic (for $\beta \neq0)$.

\section{Approximating the distribution of escape times}\label{distributionSection}

To completely characterize the escape times, we require their distribution, which captures the full range of outcomes we might expect from the stochastic dynamics of the Langevin equation. For high barriers (the Arrhenius limit) the escape is dominated by the decay of a single (quasi-equilibrium) mode in the bottom of the potential well, so the distribution is exponential with rate parameter given by $1/\tau$. On the other hand, for small barriers or sloping potentials many modes contribute and the mean may not be representative of the escape times in general. In this section, we develop an approximation to the distribution of escape times which is accurate for all $\alpha$ (i.e. any cubic potential). The approximate distribution is given as an analytical scaling form parameterized by two variables which are computed numerically for a given $\alpha$. 

\begin{figure}[t]
\includegraphics[width=\linewidth]{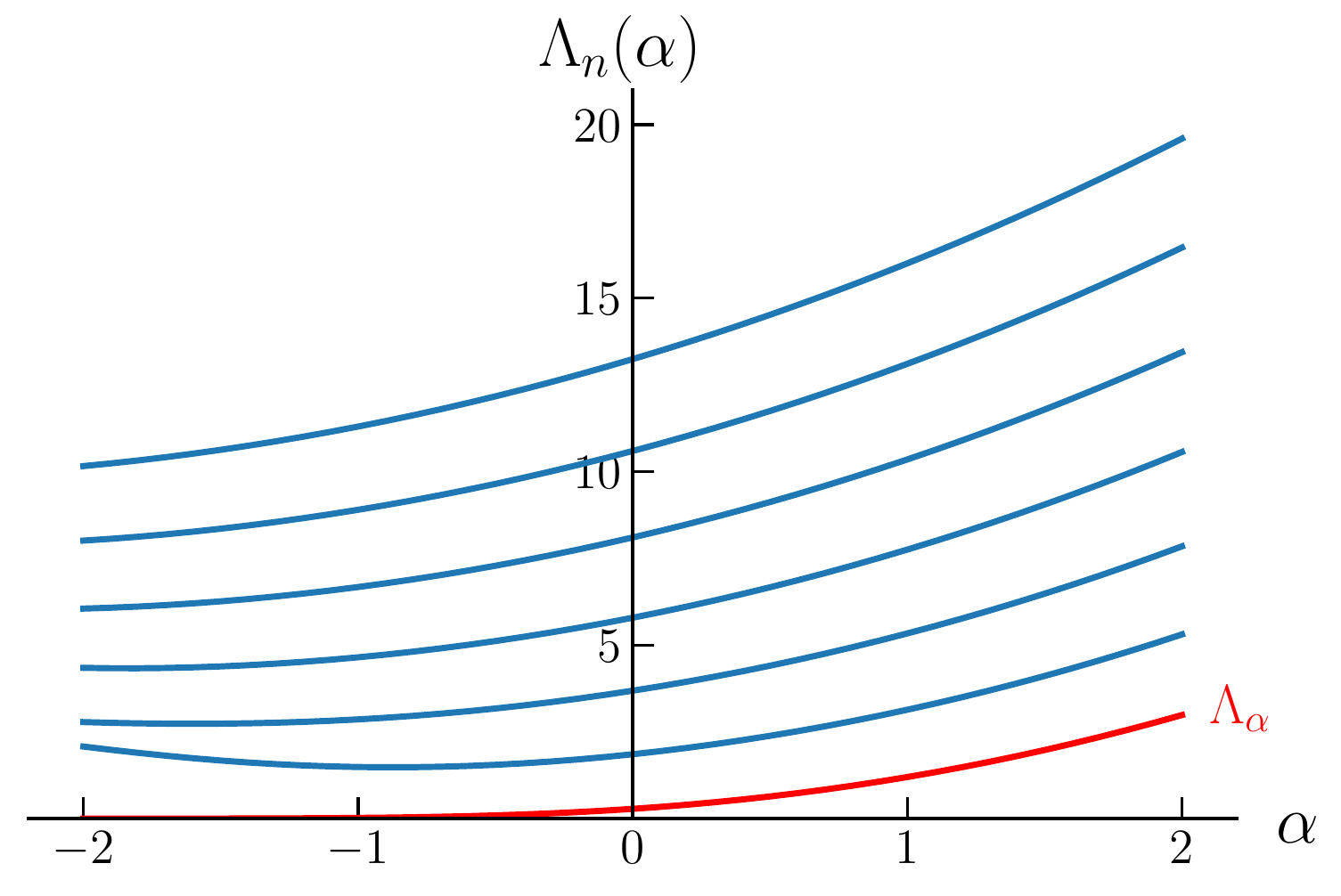}
\caption{\label{fig:eigenvalues} The scaling forms $\Lambda_n(\alpha)$ for the first seven eigenvalues. For large positive $\alpha$ the eigenvalues are approximately evenly spaced. For large negative $\alpha$ the leading eigenvalue approaches 0 and the gap to the second eigenvalue grows. The scaling form for the leading eigenvalue $\Lambda_0(\alpha) \equiv \Lambda_\alpha$ used in our approximation to the distribution of escape times is shown in red.}
\end{figure}

\subsection{First-passage distributions in Markov processes}
To study the distribution, we borrow a result from the theory of Markov processes. Consider a birth-death process, i.e. a Markov chain with a one dimensional state space and hopping only between nearest neighbor states, that has one reflecting boundary and one absorbing boundary. This is precisely what we would obtain by discretizing space in the Langevin equation Eq.~(\ref{overDampedLimit}), which has a one dimensional state space with a reflecting boundary for large negative $x$ values and an absorbing boundary for large positive $x$ values. For Markov systems initialized near the reflecting boundary, the distribution of times to reach the absorbing state can be characterized completely in terms of the eigenvalues of the transition matrix. In particular the distribution, $p(t) = -d\mathcal{P}(t)/dt$ (using the notation from the preceding section) can be shown to be a convolution of exponentials \cite{ashcroft2015mean},
\begin{equation}
p(t) = \mathcal{E}(\lambda_0) *\mathcal{E}(\lambda_1) * \cdots * \mathcal{E}(\lambda_N),
\end{equation}
where $\lambda_n$ are the negative eigenvalues of the Markov transition matrix, $\mathcal{E}(\lambda_n)$ are exponential distributions, and $*$ denotes a convolution. As we will see below, it is useful consider the Fourier transform $\tilde p(\omega)$ of the distribution $p(t)$ and the cumulant generating function
\begin{equation}\label{generatingFunction}
\log \tilde p(\omega) = \sum_{m=1}^\infty \kappa_m (i \omega)^m/m!
\end{equation}
since the cumulants $\kappa_m$ are easily expressed in terms of the eigenvalues of the Markov matrix,
\begin{equation}\label{cumulants}
\kappa_m = (m-1)! \sum_{n=0}^{N} \frac{1}{\lambda_n^m}.
\end{equation}
Note that the first cumulant $\kappa_1$ is just the mean barrier crossing time $\tau$, which was the focus of the previous section.

\subsection{Spectra of the Fokker-Planck operator}
When the state space becomes continuous, the process is generated by the Fokker-Planck operator rather than a finite transition matrix. Thus we want to understand the spectra of the right hand side of the Fokker-Planck equation,
\begin{equation}
\frac{\partial\mathcal{P}(x, t) }{\partial t}= \frac{\partial}{\partial x} \left(V'(x) \mathcal{P}(x, t) \right) + \frac{1}{2} g_0^2 \frac{\partial^2 \mathcal{P}(x,t)}{\partial x}.
\end{equation}
For numerical evaluation of the eigenvalues is convenient to change variables so that the differential operator is Hermitian. To do this we rescale $\mathcal{P}(x,t)$ by the square root of the Boltzmann factor, defining $\sigma(x,t) = \mathcal{P}(x,t)/\exp(-V(x)/g_0^2)$. Then $\sigma(x,t)$ satisfies 
\begin{equation}\label{schrodingerEq}
\frac{\partial \sigma(x,t)}{\partial t}= \left \{\frac{g_0^2}{2} \frac{\partial^2}{\partial x^2} + \frac{1}{2} V''(x) - \frac{1}{2 g_0^2} \Big(V'(x) \Big)^2 \right \} \sigma(x,t),
\end{equation}
which has the form of a Schr\"{o}dinger equation with imaginary time. Importantly, even in the limit of continuous space, the spectrum of the Fokker-Plank operator is discrete. This is easy to see from Eq.~(\ref{schrodingerEq}), because the `effective quantum potential' $(V'(x))^2/2g_0^2-V''(x)/2$ is bounded from below for the cubic potential as well as any polynomial potential that diverges super-linearly as $x\rightarrow \pm \infty$. 

Within the scaling theory developed in Section~\ref{scalingSection}, if we specialize to the cubic potential (i.e. relevant variables of the renormalization group) the eigenvalues have scaling forms
\begin{equation}
\lambda_n = g_0^{2/3} \Lambda_n(\alpha),
\end{equation}
written in terms of the RG-invariant quantity $\alpha = \epsilon/g_0^{4/3}$. This also implies a scaling form for the cumulants $\kappa_m = g_0^{-2 m/3} \mathcal{K}_m(\alpha)$. The scaling forms $\Lambda_n(\alpha)$ are plotted in Fig.~(\ref{fig:eigenvalues}) for the first several eigenvalues. These curves describing the $\alpha$-dependence of the Fokker-Planck spectra are universal for systems near a saddle node bifurcation (where $\alpha$ is the only relevant variable). For large positive $\alpha$, the eigenvalues are approximately evenly spaced, while for large negative $\alpha$, a single slowly decaying mode with eigenvalue very close to 0 dominates the behavior.

\subsection{Evenly spaced eigenvalue approximation}

To develop an approximation to the distribution of barrier escape times for the cubic potential, we assume the eigenvalues are equally spaced, $\Lambda_n(\alpha) = \Lambda_\alpha + n \Delta_\alpha$, where $\Lambda_\alpha \equiv \Lambda_0(\alpha)$ is the decay rate of the slowest decaying eigenmode. This approximation becomes exact in the limit $\alpha \rightarrow \infty$, for which the effective potential in Eq.~(\ref{schrodingerEq}) becomes harmonic. Though we can see in Fig.~(\ref{fig:eigenvalues}) that this approximation is clearly not correct for $\alpha \rightarrow -\infty$, the behavior in this limit is dominated by the vanishing leading eigenvalue $\Lambda_\alpha$. For instance, the cumulants in Eq.~(\ref{cumulants}) are insensitive to $\Delta_\alpha$ when $\Delta_\alpha \gg \Lambda_\alpha$ and $\Lambda_\alpha \rightarrow 0$. Thus, this approximation will produce a family of distributions which correctly captures the behavior for both large negative and positive $\alpha$. As we will see below, the approximation is also quite accurate over the full range of $\alpha$. 

One caveat of the equal spacing approximation is that the mean escape time $\mathcal{T} = \mathcal{K}_0 = \sum_{n=1}^\infty \Lambda_n^{-1}$ always diverges. Therefore, we must fix the mean to the value derived in Section~\ref{meanTime}. On the other hand, the higher order cumulants, which determine the shape of the distribution, always converge.

To proceed, we sum Eq.~(\ref{cumulants}) using our eigenvalue ansatz. For $m\geq2$ the result is,
\begin{equation}
\mathcal{K}_m(\alpha) = (-1)^m \frac{\psi^{(m-1)}(\Lambda_\alpha/\Delta_\alpha)}{\Delta_\alpha^m},
\end{equation}
where $\psi^{(m)}(x)$ is the polygamma function of order $m$. Neglecting the mean, we can sum the series Eq.~(\ref{generatingFunction}), then exponentiate and inverse Fourier transform to obtain the distribution $p(t)$. Writing the distribution in terms of its universal scaling function, $p(t,\epsilon, g_0) = g_0^{2/3} \rho(s, \alpha)$ with $s = (t- \tau)/g_0^{2/3}$ and $\alpha = \epsilon/g_0^{4/3}$, the final result is
\begin{equation}\label{approxDistribution}
\begin{split}
\rho(s, \alpha) = \frac{\Delta_\alpha}{\Gamma(\Lambda_\alpha/\Delta_\alpha)} \exp \bigg( &e^{\psi^{(0)}(\Lambda_\alpha/\Delta_\alpha) - \Delta_\alpha \, s }  - \Lambda_\alpha s  \\
&\quad +  \frac{\Lambda_\alpha}{\Delta_\alpha} \psi^{(0)}(\Lambda_\alpha/\Delta_\alpha) \bigg).
\end{split}
\end{equation}
To use this family of distributions, we compute the leading eigenvalue $\Lambda_\alpha$ using a shooting method on the right hand side of Eq.~(\ref{schrodingerEq}). The effective $\Delta_\alpha$ is chosen so that the second cumulant $\mathcal{K}_2$ (i.e. the variance) is exact. To do this, we follow the same approach used in Section~\ref{meanTime} to write an integral expression for the variance. The result is $\mathcal{K}_2= \tau_2 - \tau^2$ with
\begin{equation}
 \tau_2(x_i | x_f) = \frac{2}{g_0^2}\int_{x_i}^{x_f} dy \int_{-\infty}^y dz \, \tau(z | x_f) \, e^{-\frac{2}{g_0^2} \left[V(z) - V(y) \right]},
\end{equation}
where $\tau(z | x_f)$ is the mean given in Eq.~(\ref{fptFunctional}). After evaluating these integrals numerically to obtain the variance, we solve $\mathcal{K}_2 = \psi^{(1)}(\Lambda_\alpha)/\Delta_\alpha^2$ to fix the eigenvalue spacing $\Delta_\alpha$.

\begin{figure*}[t]
\includegraphics[width=0.49\linewidth]{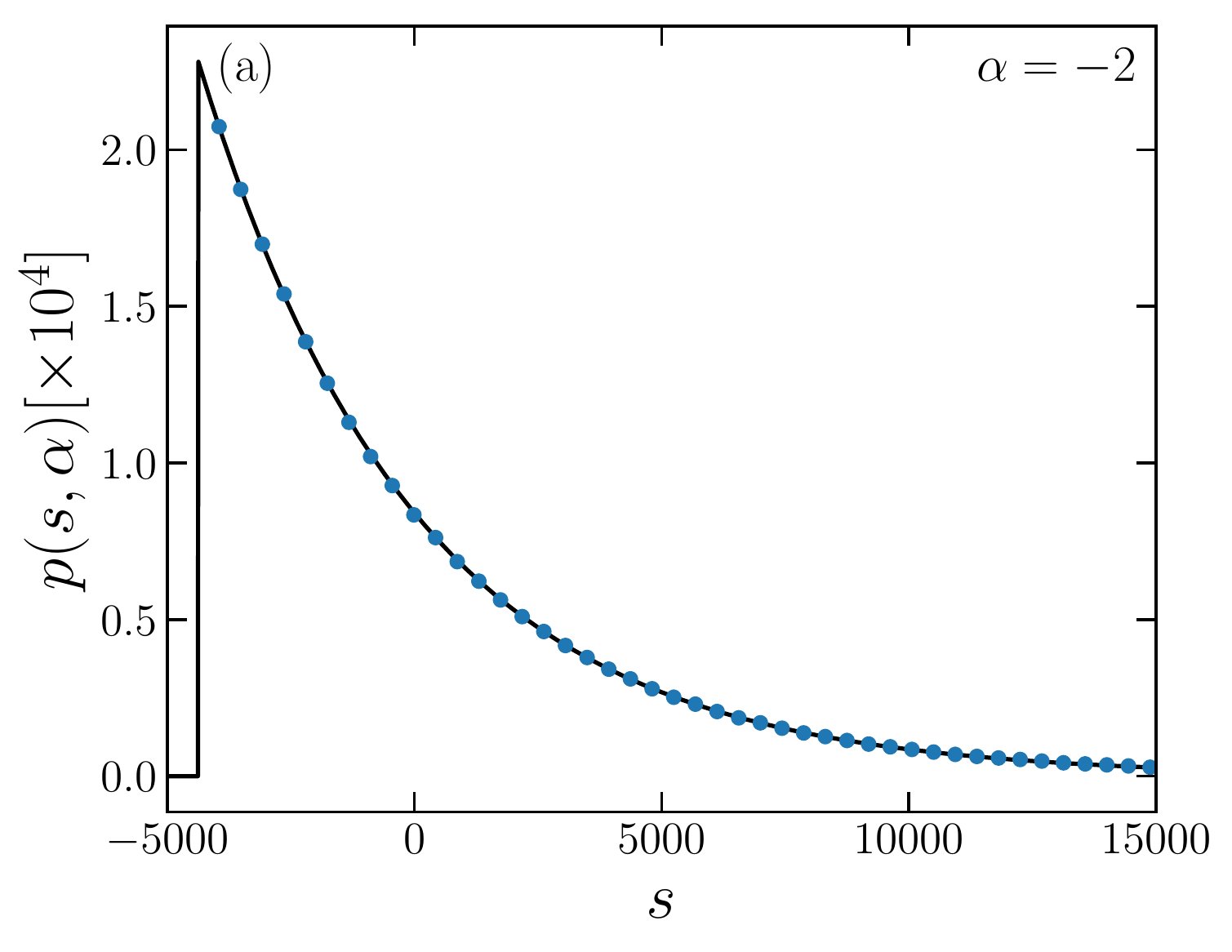} \,\,\,
\includegraphics[width=0.49 \linewidth]{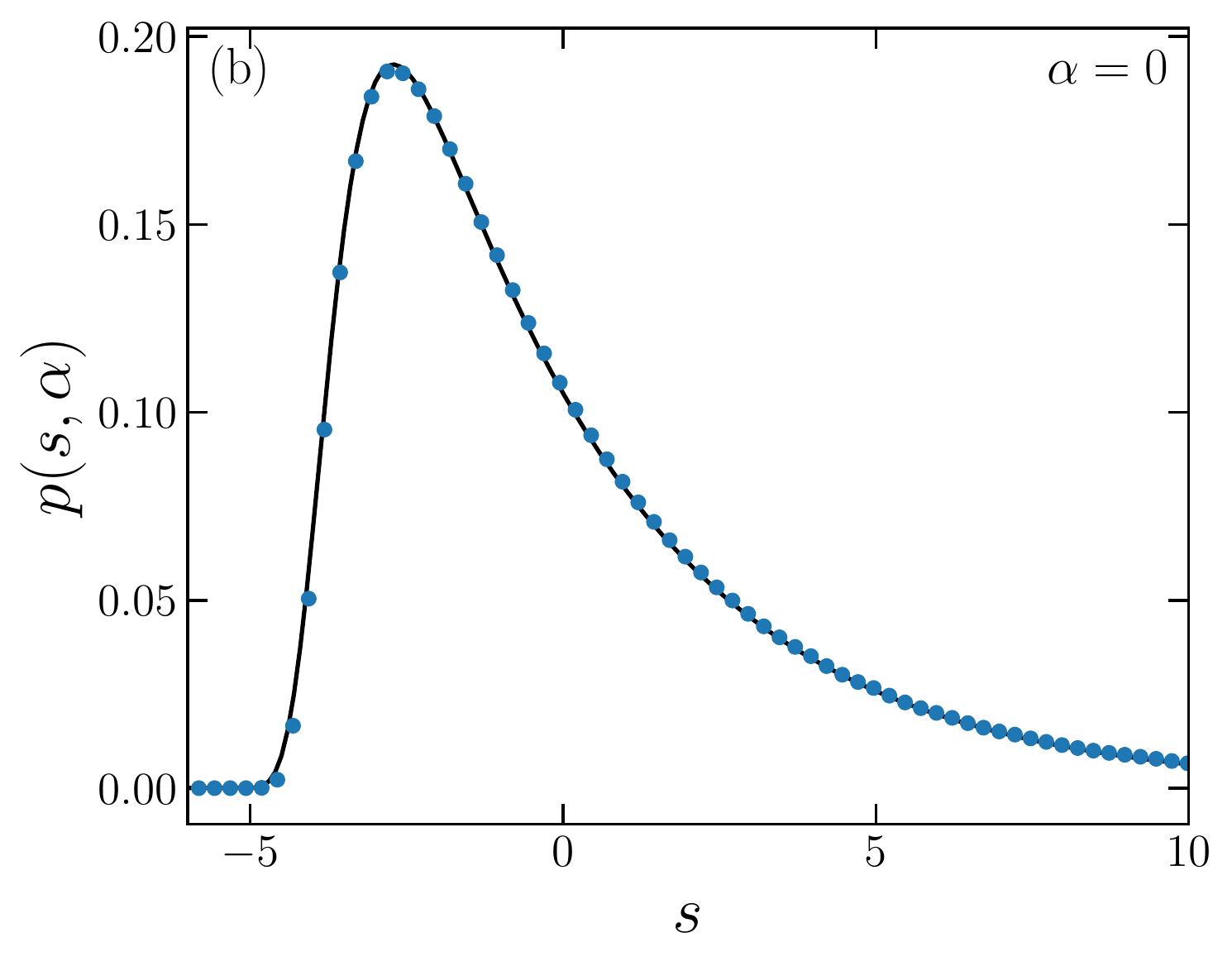} \\  \vspace{0.2em}
\includegraphics[width=0.49 \linewidth]{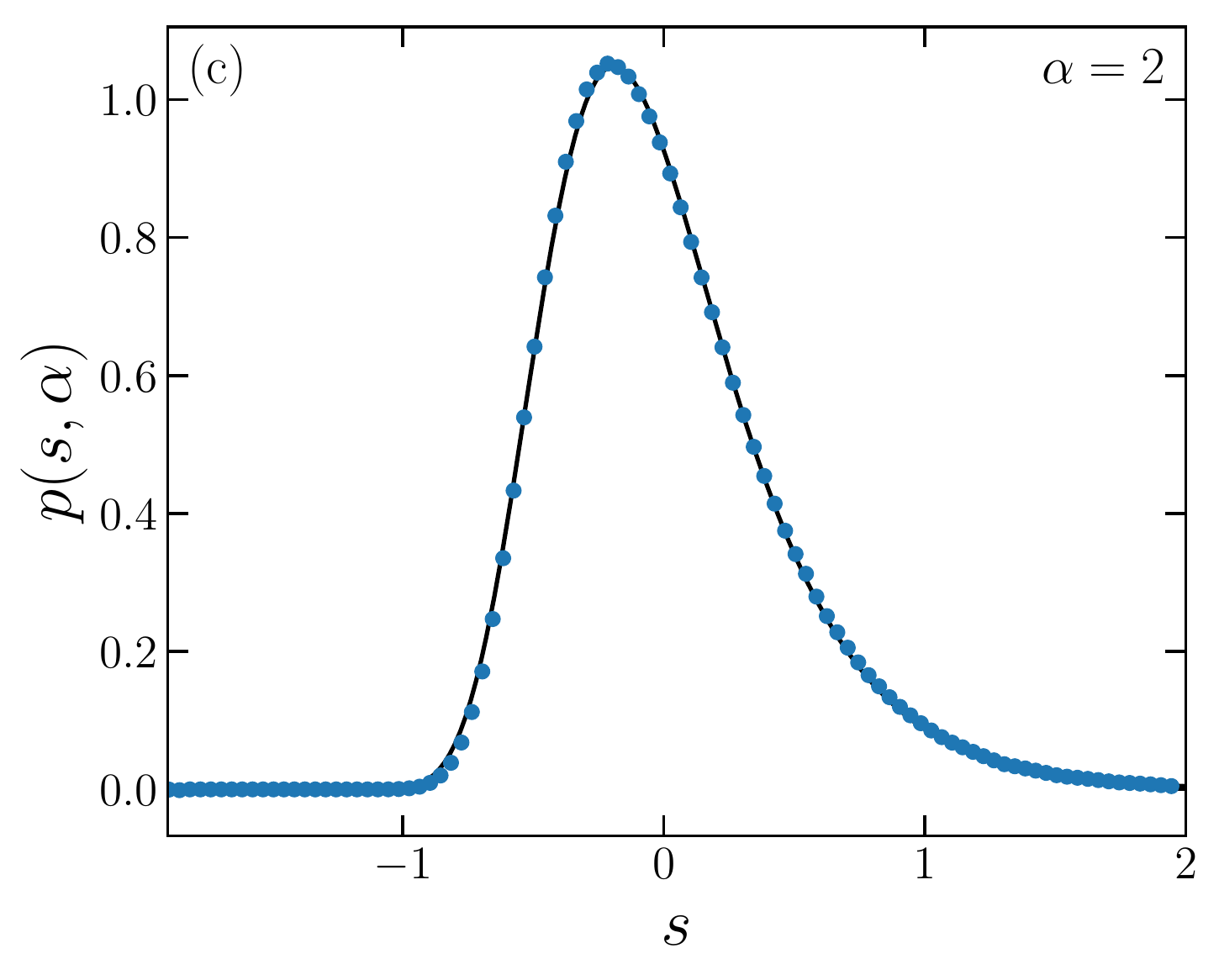} \,\,\,
\includegraphics[width=0.49 \linewidth]{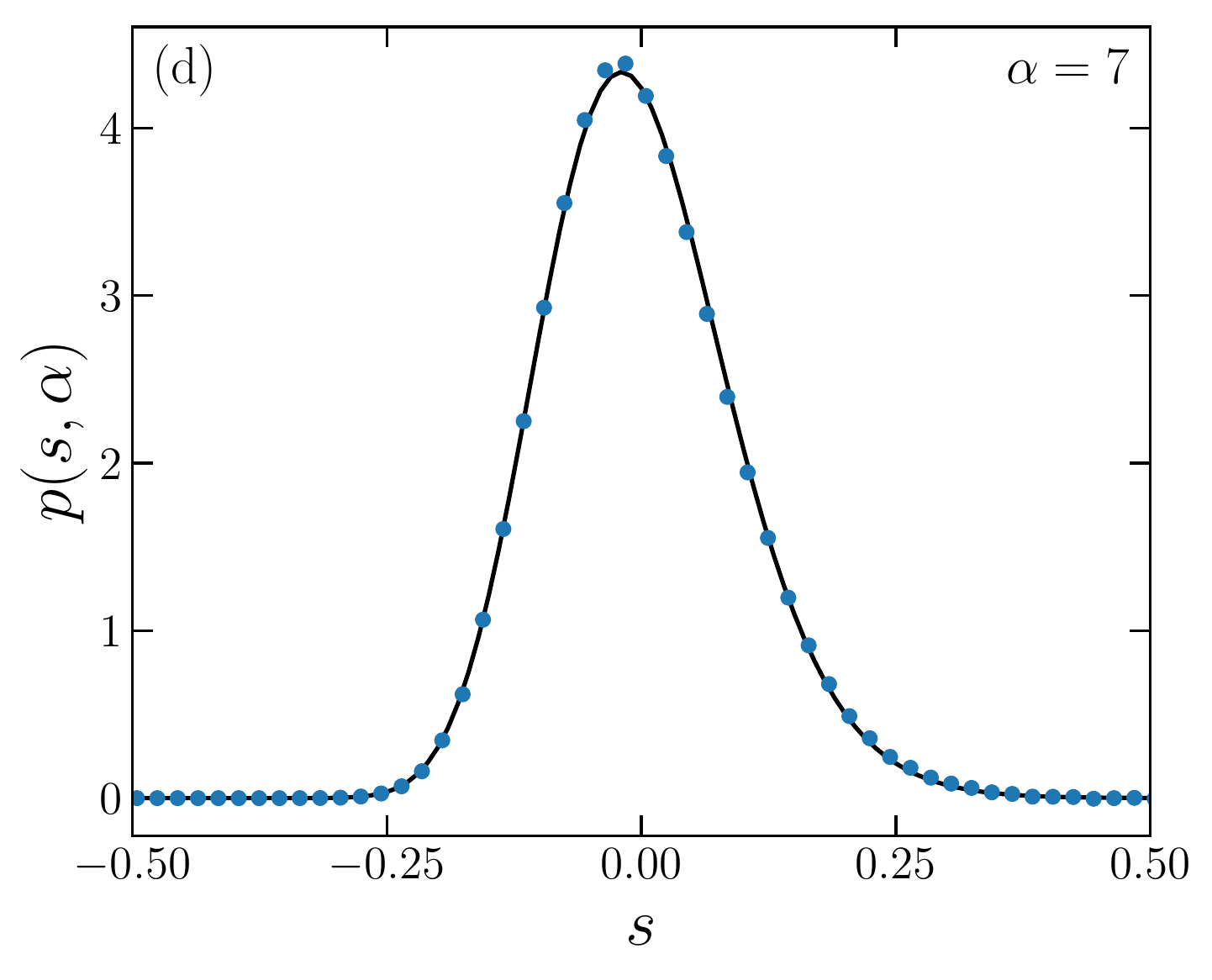}
\caption{\label{fig:distributions} The barrier crossing time distributions obtained using our evenly spaced eigenvalue approximation Eq.~(\ref{approxDistribution}) (lines) and from direct simulation of the Fokker-Planck equation (symbols) for (a) $\alpha = -2$, (b) $\alpha = 0$, (c) $\alpha = 2$, and (d) $\alpha = 7$. In all cases agreement between the theory and simulations is excellent. In the large barrier limit (a) the distribution is approximately exponential and in the strongly sloped potential (d) it is nearly Gaussian.}
\end{figure*}

In Figure~\ref{fig:distributions} we compare the approximate distribution Eq.~(\ref{approxDistribution}) to those obtained by direct numerical simulation of the Fokker-Plank equation. Our approximation captures the shape of the distributions remarkably well for the full range of $\alpha$. 

We can also show analytically that this distribution reproduces the correct large $\alpha$ limits. For large negative $\alpha$, taking the $\Delta_\alpha \rightarrow \infty$ limit of Eq.~(\ref{approxDistribution}), gives the expected exponential distribution, $\rho(s, \alpha) \to \Lambda_\alpha e^{-s \Lambda_\alpha - 1}$ (shifted to have zero mean). For large positive we can estimate $\Lambda_\alpha$ and $\Delta_\alpha$ using the harmonic approximation to the potential in Eq.~(\ref{schrodingerEq}). We find $\Lambda_\alpha \sim \alpha^2$ and $\Delta_\alpha \sim \sqrt{\alpha}$ so that the cumulants scale like $\mathcal{K}_m(\alpha) \sim \alpha^{3(1-m)/2}$ for $\alpha \to \infty$. In particular, the ratio $\mathcal{K}_m/\mathcal{K}_2^{m/2} \to 0$ for $m>2$, i.e. the higher order cumulants are small compared to the variance and the distribution approaches a Gaussian for large $\alpha$. This prediction is born out in direct simulation of the Fokker-Plank equation (see Figure~\ref{fig:distributions}d).

The distribution Eq.~(\ref{approxDistribution}) combined with our analytical understanding of the mean escape time provides a complete description of the barrier crossing process. Our procedure allows for accurate approximations to the distribution by evaluating just the two universal scaling forms $\Lambda_\alpha$ and $\Delta_\alpha$. Knowledge of these quantities allows for evaluation of the distribution of barrier crossing times for any system near a saddle-node bifurcation.

\section{Discussion}
\label{discussionSection}

We expect our results will be directly applicable to barrier crossing processes in which thermal fluctuations are comparable to the energy barrier including the aforementioned experimental systems, narrow escape problems in cellular biology \cite{schuss2007narrow}, and downhill protein folding scenarios \cite{sabelko1999observation, best2006diffusive}. Our approximation to the distribution of barrier escape times, combined with our analytical results for the mean, provides an accurate and complete characterization of the barrier crossing process for systems near a saddle node bifurcation. 

A more thorough analysis of incorporating perturbative corrections from irrelevant variables into Eq.~(\ref{fptFunctional}) would be both theoretically interesting and useful in applications. The interplay between anharmonic corrections and finite initial conditions is also important. Computing these corrections will extend the applicability of our theory to systems in which the boundary conditions correspond to positions with non-negligible anharmonicity. It will be interesting to test the accuracy of our evenly spaced eigenvalue approximation to the barrier escape time distribution when irrelevant variables are incorporated. We conjecture that, at least perturbatively, Eq.~(\ref{approxDistribution}) will still accurately parameterize the distributions if $\Lambda_\alpha$ and $\Delta_\alpha$ are corrected to account for the irrelevant variables.

Our analysis directly translates to higher order cuspoid catastrophes \cite[Section 36.2]{NIST:DLMF}, which form their own universality classes with different exponents (in fact, these have already been analyzed for the discrete iterated map  \cite{hirsch1982intermittencyRG, hu1982exact}). For these bifurcations, the fixed point potential will be a higher order monomial and our analysis can be used to identify the relevant variables and develop a scaling theory for quantities like the barrier crossing time mean and distribution.

More generally, it would be useful to study the applicability of our renormalization group and scaling analysis to systems with colored noise, multiple dimensions, or in other damping regimes. The effects of colored noise are encoded in the correlation function $\langle \xi(t) \xi(t') \rangle =G(x, t-t')$. The renormalization group transformation can be adapted to act on the Fourier transform of this quantity $\tilde{G}(x, \omega)$, giving flows of the colored noise under coarse-graining
(for example, barrier crossing between two symmetric wells 
-- a noisy pitchfork bifurcation -- coupled to an Ohmic heat bath leads
in the quantum limit to a critical point in the same universality class
as the Kondo problem~\cite{Chakravarty82}).
We expect short-range correlations will be irrelevant under coarse-graining, while those with power-law decay will give rise to new anomalous scaling.  
For some reactions, an underdamped model or multi-dimensional reaction coordinate may be required for an accurate description. Renormalization group scaling will provide a natural organizing framework for these studies.

Nucleation of abrupt phase transitions (e.g., raindrop formation) is also 
described by Arrhenius rates. Here the noiseless bifurcation underlying
critical droplet theory is the {\em spinodal line}. This line -- a
discredited mean-field boundary between nucleation and spontaneous phase
separation -- could play the role of our renormalization-group fixed point
in a future generalization of this work to higher spatial dimensions.

Finally, the saddle node bifurcation is the simplest example of a depinning transition. One anticipates studying how adding noise would affect depinning of earthquakes, vortices in superconductors, plastic flow in crystals, raindrops on windshields, coffee soaking into napkins, and other depinning phenomena. Each of these has anomalous exponents and RG treatments~\cite{fisher1998collective} even without added noise. Our work provides a stepping stone toward understanding the universal scaling near noisy depinning transitions in these more sophisticated systems as well.

\bibliography{references}

\end{document}